\documentclass[english]{article}%
\usepackage[T1]{fontenc}
\usepackage[latin9]{inputenc}
\usepackage{verbatim}
\usepackage{float}
\usepackage{amsmath}
\usepackage{amssymb}
\usepackage{graphicx}
\usepackage{amssymb}
\usepackage{latexsym}
\usepackage{epsfig}
\usepackage{float}
\usepackage{dsfont}
\usepackage{babel}
\usepackage{amsfonts}%
\makeatletter

\newcommand{\be}{\begin{equation}}
\newcommand{\ee}{\end{equation}}

\allowdisplaybreaks
\makeatother
\begin{document}
{}~ \hfill\vbox{\hbox{CTP-SCU/2018004}}
\break\vskip 3.0cm

\begin{center}
\Large \bf
Temperature Dependence of In-plane Resistivity and\\
\vspace*{2ex}
Inverse Hall Angle in NLED Holographic Model
\end{center}

\vspace*{10ex}
\centerline{\large Qingyu Gan, Peng Wang, Haitang Yang}
\vspace*{7ex}
\vspace*{4ex}
\centerline{\large \it Center for theoretical physics}
\centerline{\large \it Sichuan University}
\centerline{\large \it Chengdu, 610064, China}
\vspace*{1ex}
\vspace*{4ex}

\centerline{2016222020006@stu.scu.edu.cn, pengw@scu.edu.cn, hyanga@scu.edu.cn}
\vspace*{10ex}
\centerline{\bf Abstract}
\bigskip
\smallskip

In the strange metal phase of the high-$T_{c}$ cuprates, it is challenging to explain the linear temperature dependence of the in-plane resistivity and the quadratic temperature dependence of the inverse Hall angle. In this paper, we investigate the temperature dependence of the in-plane resistivity and inverse Hall angle in the nonlinear electrodynamics holographic model developed in our recent work. Maxwell electrodynamics and Born-Infeld electrodynamics are considered. Both cases support a wide spectrum of temperature scalings in parameter space. For Maxwell electrodynamics, the T-linear in-plane resistivity generally dominates at low temperatures and survives into higher temperatures in a narrow strip-like manner. Meanwhile, the T-quadratic inverse Hall angle dominates at high temperatures and extends down to lower temperatures. The overlap between the T-linear in-plane resistivity and the T-quadratic inverse Hall angle, if occurs, would generally present in the intermediate temperate regime. The Born-Infeld case with $a>0$ is quite similar to the Maxwell case. For the Born-Infeld case with $a<0$, there can be a constraint on the charge density and magnetic field. Moreover, the overlap can occur for strong charge density.

\vfill
\eject
\baselineskip=16pt \vspace*{10.0ex}
%\tableofcontents

\section{Introduction}

The holographic techniques developed within string theory \cite{L.Susskind, Maldacena, E.Witten} have been widely applied in condensed matter
physics to explore the properties of strongly correlated systems and undergone
some remarkable progress \cite{Gubser,Karch:2007pd,S. A. Hartnoll,S. S. Lee,H. Liu,M. Cubrovic,Hartnoll:2009sz,Herzog:2009xv,S. Sachdev}.

In present of the strong interactions, the transport properties behave
unconventionally in contrast to normal materials which is described by Fermi
liquid theorem. A prime example is the strange metal phase emerging from the
normal states of the high-$T_{c}$ cuprates. Among a number of weird transport
properties, two of them are celebrated: one is the linear temperature
dependence of the in-plane resistivity $R_{xx}$ and the other is the quadratic
temperature dependence of the inverse Hall angle $\cot\Theta_{H}$ \cite{S.
Martin,Ando,R. A. Cooper et al., Chien, Tyler}. A comprehensive review on this
aspect could be found in \cite{Hussey}. As early as 90s, some theoretical
attempts have been made to explain these behaviors. It is suggested that the
scattering of spinons and holons are governed by two different relaxation
times in \cite{P. W. Anderson}, and the independent behaviors of charge
conjugation odd and even quasi-particles is proposed in \cite{P. Coleman}.

In recent decade, many holographic models are developed to illustrate and
realize these anomalous scalings.{} The T-linear resistivity is reproduced in
\cite{Hartnoll:2009ns,Davison:2013txa} by holography. In \cite{Mike Blake}, it
claimed that charge-conjugation symmetric conductivity does not contribute to
the inverse Hall angle within an Einstein-Maxwell-Dilaton (EMD) model. More
EMD-like theories can be found in
\cite{Blake:2015ina,Amoretti:2015gna,Zhou:2015dha,Ge:2016sel,Cremonini:2016avj}
. In \cite{Erin Blauvelt}, it realized the T-linear
resistivity and T-quadratic inverse Hall angle simultaneously in a nonlinear
Dirac-Born-Infeld (DBI) model. More DBI-like theories can be found in
\cite{Pal:2010sx,Bom Soo Kim,Karch:2014mba,Kiritsis:2016cpm,Cremonini:2017qwq,Kuang:2018ymh}.

Our previous work \cite{Wang:2018hwg} considered a  holographic model with a generic
nonlinear electrodynamic (NLED) field. Our NLED holographic model has been taken into account the full backreaction and momentum dissipation following the methods in \cite{Cremonini:2017qwq} and \cite{Andrade}, respectively. In this paper, We follow up our recent work to
investigate the scalings of temperature dependence of $R_{xx}$ and $\cot
\Theta_{H}$ in parameter space of this model. Furthermore, we investigate if
there is overlap between the T-linear $R_{xx}$ and T-quadratic $\cot\Theta
_{H}$. To extract the effective scalings of temperature, we take advantage of
the density plots of $d\log_{10}(dR_{xx}/dT)/d\log_{10}T$ and $d\log
_{10}(d\cot\Theta_{H}/dT)/d\log_{10}T$ in parameter space.

The rest of this article is organized as follows. In section
\ref{sec:Holographic-Setup-and}, we briefly review the holographic model set
up in \cite{Wang:2018hwg} and give the expressions for in-plane resistivity
$R_{xx}$ and inverse Hall angle $\cot\Theta_{H}$. In section
\ref{sec:In-Plane-transport-properties}, we show the density plots of
$d\log_{10}(dR_{xx}/dT)/d\log_{10}T$ and $d\log_{10}(d\cot\Theta_{H}
/dT)/d\log_{10}T$ in various parameter spaces to study the scalings of
temperature dependence of $R_{xx}$ and $\cot\Theta_{H}$. We focus on two
typical cases: one is Maxwell electrodynamics and the other is nonlinear
Born-Infeld electrodynamics which are discussed in subsections
\ref{subsec:Maxwell-Electrodynamics} and
\ref{subsec:Born-Infeld-Electrodynamics}, respectively. In section
\ref{sec:Discussion-and-conclusion}, we end in short conclusions and discussions.

\section{Holographic Setup and DC Conductivity}

\label{sec:Holographic-Setup-and}

Consider a holographic model with the action given by
\begin{equation}
S=\int d^{4}x\sqrt{-g}[R-2\Lambda-\frac{1}{2}\sum_{I=1}^{2}(\partial\psi
_{I})^{2}+\mathcal{L}(s,p)], \label{eq:action}%
\end{equation}
where $\Lambda=-3/l^{2}$, and we take $16\pi G=1$ and $l=1$ for simplicity. To
break translational symmetry and generate momentum dissipation, we introduce
two axions $\psi_{I}$ $(I=1,2)$, which lead to a finite DC conductivity
\cite{Andrade,M. Blake}. In the action $\left(  \ref{eq:action}\right)  $, the
generic NLED Lagrangian $\mathcal{L}(s,p)$ is a function of $s$ and
$p$, where $s\equiv-F^{ab}F_{ab}/4$ and $p\equiv-\epsilon^{abcd}F_{ab}%
F_{cd}/8$ (the indices $a,b\cdots$ denote the bulk spacetime $t$, $r$, $x$ and
$y$). The two independent nontrivial scalars $s$ and $p$ are built from the
electromagnetic field $A_{a}$ using field strength tensor $F_{ab}=\partial
_{a}A_{b}-\partial_{b}A_{a}$ and totally anti-symmetric Lorentz tensor
$\epsilon^{abcd}$.

Along the lines of \cite{Wang:2018hwg}, we take the following ansatz to
construct a black brane solution with asymptotic AdS spacetime:
\begin{align}
ds^{2}  &  =-f(r)dt^{2}+\frac{dr^{2}}{f(r)}+r^{2}(dx^{2}+dy^{2}),\nonumber\\
\mathbf{A}  &  =A_{t}(r)dt+\frac{h}{2}(xdy-ydx),\nonumber\\
\psi_{1}  &  =\alpha x,\label{eq:black brane solution}\\
\psi_{2}  &  =\alpha y,\nonumber
\end{align}
where $h$ denotes the magnetic field and $\alpha$ denotes the strength of
momentum dissipation. Plugging the ansatz into the action $(\ref{eq:action})$
and varying it with respect to $g_{ab}$, $A_{a}$ and $\psi_{I}$, we obtain the
equations of motions:
\begin{align}
f(r)-3r^{2}+rf^{^{\prime}}(r)  &  =-\frac{\alpha^{2}}{2}+\frac{r^{2}}%
{2}[\mathcal{L}(s,p)+A_{t}^{^{\prime}}(r)G^{rt}],\label{eq:rh equation}\\
2f^{^{\prime}}(r)-6r+rf^{^{\prime\prime}}(r)  &  =r[\mathcal{L}(s,p)+hG^{xy}%
],\\
{}[r^{2}G^{rt}]^{^{\prime}}  &  =0, \label{eq:G equation}%
\end{align}
where the prime denotes the derivative to radial direction, and $G$ is defined
as $G^{ab}\equiv-\partial\mathcal{L}(s,p)/\partial F_{ab}$. Eqn.
$(\ref{eq:G equation})$ leads to $G^{rt}=-\rho/r^{2},$ where $\rho$ could be
interpreted as the charge density of the dual field theory \cite{Wang:2018hwg}. The horizon located at $r_{h}$ is determined by $f(r_{h})=0$, and the
Hawking temperature is given by $T=f^{^{\prime}}(r_{h})/4\pi$. Therefore, eqn.
$(\ref{eq:rh equation})$ gives
\begin{equation}
-3r_{h}^{2}+4\pi r_{h}T=-\frac{\alpha^{2}}{2}+\frac{r_{h}^{2}}{2}%
[\mathcal{L}(s_{h},p_{h})+A_{t}^{^{\prime}}(r_{h})G_{h}^{rt}] \label{eq:rh}%
\end{equation}
with
\begin{align}
s_{h}  &  =\frac{A_{t}^{^{\prime}2}(r_{h})}{2}-\frac{h^{2}}{2r_{h}^{4}%
},\nonumber\\
p_{h}  &  =-\frac{hA_{t}^{^{\prime}}(r_{h})}{r_{h}^{2}}%
,\label{eq:relations between}\\
G_{h}^{rt}  &  =-\frac{\rho}{r_{h}^{2}}=-\mathcal{L}^{(1,0)}(s_{h},p_{h}%
)A_{t}^{^{\prime}}(r_{h})+\mathcal{L}^{(0,1)}(s_{h},p_{h})\frac{h}{r_{h}^{2}%
},\nonumber
\end{align}
where the superscripts $(1,0)$ and $(0,1)$ denote the partial derivative of
$\mathcal{L}(s,p)$ with respect to $s$ and $p$, respectively.

Via gauge/gravity duality, the electromagnetic field $A_{a}$ living in the
bulk would be dual to a conserved current $\mathcal{J}^{i}$ (the indices
$i,j\cdots$ denote the co-dimensional boundary spacetime $t$, $x$ and $y$)
living in the boundary. As a consequence, the DC conductivities for
$\mathcal{J}^{i}$ can be derived using the method developed in \cite{Mike
Blake,Donos:2014uba}. The detailed calculation was carried out in
\cite{Wang:2018hwg}, and here we only display the final expressions for DC
conductivities $\sigma$:
\begin{align}
\sigma_{xx}  &  =\frac{\alpha^{2}r_{h}^{2}\left[  h^{2}+\frac{\alpha^{2}%
r_{h}^{2}}{\mathcal{L}^{(1,0)}(s_{h},p_{h})}+A_{t}^{^{\prime}2}(r_{h}%
)r_{h}^{4}\right]  }{\left[  h^{2}+\frac{\alpha^{2}r_{h}^{2}}{\mathcal{L}%
^{(1,0)}(s_{h},p_{h})}\right]  ^{2}+h^{2}A_{t}^{^{\prime}2}(r_{h})r_{h}^{4}%
},\nonumber\\
\sigma_{xy}  &  =\frac{hA_{t}^{^{\prime}}(r_{h})r_{h}^{2}\left[  2\alpha
^{2}r_{h}^{2}+\mathcal{L}^{(1,0)}(s_{h},p_{h})(h^{2}+A_{t}^{^{\prime}2}%
(r_{h})r_{h}^{4})\right]  }{\left[  h^{2}+\frac{\alpha^{2}r_{h}^{2}%
}{\mathcal{L}^{(1,0)}(s_{h},p_{h})}\right]  ^{2}+h^{2}A_{t}^{^{\prime}2}%
(r_{h})r_{h}^{4}}-\mathcal{L}^{(0,1)}(s_{h},p_{h}).
\label{eq:DC conductivities}%
\end{align}
To express DC conductivities in terms of the temperature $T$, the charge
density $\rho,$ the magnetic field $h$, the strength of momentum dissipation
$\alpha$, we need to solve eqns. $(\ref{eq:rh equation})$ and
$(\ref{eq:G equation})$ for $r_{h}$ and $A_{t}^{^{\prime}}(r_{h})$ and plug
them into eqns. $(\ref{eq:DC conductivities})$. The in-plane resistivity
$R_{xx}$ and inverse Hall angle $\cot\Theta_{H}$ are defined as
\begin{equation}
R_{xx}=\frac{\sigma_{xx}}{\sigma_{xx}^{2}+\sigma_{xy}^{2}}\text{ and }%
\cot\Theta_{H}=\frac{\sigma_{xx}}{\sigma_{xy}}.
\label{eq:resistivity and Hall angle}%
\end{equation}
Notice that $R_{xx}$ and $\cot\Theta_{H}$ remain invariant under the scaling
transformation
\begin{equation}
T\rightarrow\lambda T,\alpha\rightarrow\lambda\alpha,h\rightarrow\lambda
^{2}h,\rho\rightarrow\lambda^{2}\rho\label{eq:scaling}%
\end{equation}
for some positive constant $\lambda.$ From now on we will rescale $T$, $h$,
$\rho$ and $\alpha$ to $T/\alpha$, $h/\alpha^{2}$, $\rho$/$\alpha^{2}$ and $1$
by taking the scaling factor $\lambda=1/\alpha$.

\section{Temperature Dependence of $R_{xx}$ and $\cot\Theta_{H}$}

\label{sec:In-Plane-transport-properties}

In this section we will discuss the scalings of the temperature dependence of
the in-plane resistivity $R_{xx}$ and inverse Hall angle $\cot\Theta_{H}$ in
parameter space spanned by $T/\alpha$, $h/\alpha^{2}$, $\rho/\alpha^{2}$ and
some possible parameters from $\mathcal{L}(s,p)$. However, the temperature
dependence of $R_{xx}$ and $\cot\Theta_{H}$ is highly nontrivial. To compare
with the results from experiments, we can fit $R_{xx}$ and $\cot\Theta_{H}$
with some power of $T/\alpha$:
\begin{align}
R_{xx}  &  \sim A+B\left(  T/\alpha\right)  ^{n},\nonumber\\
\cot\Theta_{H}  &  \sim C+D\left(  T/\alpha\right)  ^{m},
\label{eq:experimence}%
\end{align}
where the terms $A$, $B$, $C$ and $D$ can depend on $h/\alpha^{2}$ and
$\rho/\alpha^{2}$. The effective power factors $n$ and $m$ are usually focused
in experiments since they mainly govern the temperature dependence of $R_{xx}$
and $\cot\Theta_{H}$. To extract the effective power factors $n$ and $m$ from
eqn. $\left(  \ref{eq:experimence}\right)  $, we display the density plots of
$d\log_{10}(dR_{xx}/dT)/d\log_{10}T$ and $d\log_{10}(d\cot\Theta_{H}%
/dT)/d\log_{10}T$ in the parameter space.{} For latter convenience, we
introduce $N$ and $M$ as
\begin{align}
\frac{d\log_{10}(dR_{xx}/dT)}{d\log_{10}T}\equiv N  &  \Longrightarrow
R_{xx}\sim(T/\alpha)^{N+1},\label{eq:notation R}\\
\frac{d\log_{10}(d\cot\Theta_{H}/dT)}{d\log_{10}T}\equiv M  &  \Longrightarrow
\cot\Theta_{H}\sim(T/\alpha)^{M+1}, \label{eq:notation C}%
\end{align}
where $N=0$ and $M=1$ correspond to the linear temperature dependence of
in-plane resistivity $R_{xx}$ and the quadratic temperature dependence of
inverse Hall angle $\cot\Theta_{H}$, respectively. In the following, we focus
on Maxwell electrodynamics in subsection \ref{subsec:Maxwell-Electrodynamics}
and Born-Infeld electrodynamics in subsection
\ref{subsec:Born-Infeld-Electrodynamics}.

\subsection{Maxwell Electrodynamics}

\label{subsec:Maxwell-Electrodynamics}

The Lagrangian for Maxwell Electrodynamics reads
\begin{equation}
\mathcal{L}(s,p)=s\text{.} \label{eq:Mawell Lagrange}%
\end{equation}
Combining eqns. $(\ref{eq:rh})$, $(\ref{eq:relations between})$,
$(\ref{eq:DC conductivities})$, $(\ref{eq:resistivity and Hall angle})$ and
$(\ref{eq:Mawell Lagrange})$, one can obtain
\begin{align}
R_{xx}  &  =\frac{r_{h}^{2}\alpha^{2}(h^{2}+\rho^{2}+r_{h}^{2}\alpha^{2}%
)}{r_{h}^{4}\alpha^{4}+2r_{h}^{2}\alpha^{2}\rho^{2}+h^{2}\rho^{2}+\rho^{4}%
},\label{eq:Maxwell R}\\
\cot\Theta_{H}  &  =\frac{r_{h}^{2}\alpha^{2}(h^{2}+\rho^{2}+r_{h}^{2}%
\alpha^{2})}{h\rho(h^{2}+\rho^{2}+2r_{h}^{2}\alpha^{2})}, \label{eq:Maxwell C}%
\end{align}
with the horizon $r_{h}$ satisfying
\begin{equation}
-12r_{h}^{4}+16\pi Tr_{h}^{3}+2\alpha^{2}r_{h}^{2}+h^{2}+\rho^{2}=0.
\label{eq:Maxwell rh}%
\end{equation}
In this paper we focus on $h/\alpha^{2}\geq0$ and $\rho/\alpha^{2}\geq0$, thus
$R_{xx}$ and $\cot\Theta_{H}$ are non-negative in Maxwell electrodynamics. We
numerically solve eqn. $(\ref{eq:Maxwell rh})$ for $r_{h}$ and then use eqns.
$(\ref{eq:Maxwell R})$ and $(\ref{eq:Maxwell C})$ to study the scalings of
temperature dependence of $R_{xx}$ and $\cot\Theta_{H}$, respectively.

\subsubsection{In-plane Resistivity}

\label{subsec:Maxwell In-plane-resistivity}

\begin{figure}[ptb]
\noindent\begin{raggedright}
\includegraphics[scale=0.34]{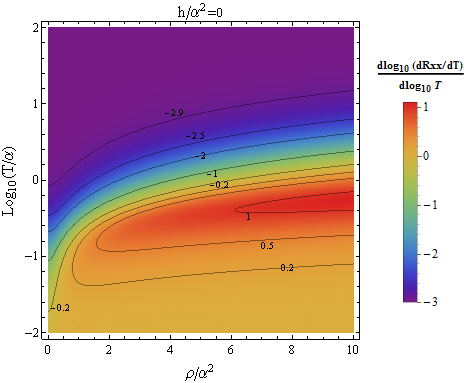}\includegraphics[scale=0.34]{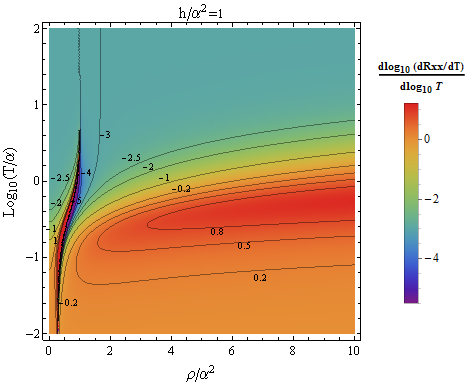}\includegraphics[scale=0.34]{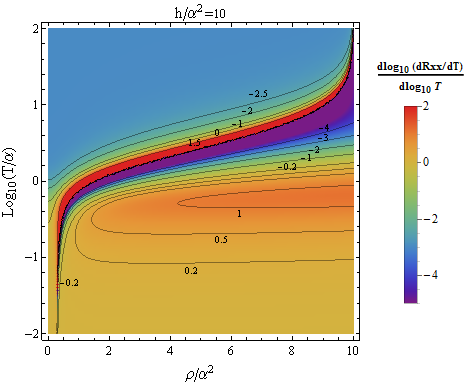}
\par\end{raggedright}
\noindent\begin{raggedright}
\includegraphics[scale=0.34]{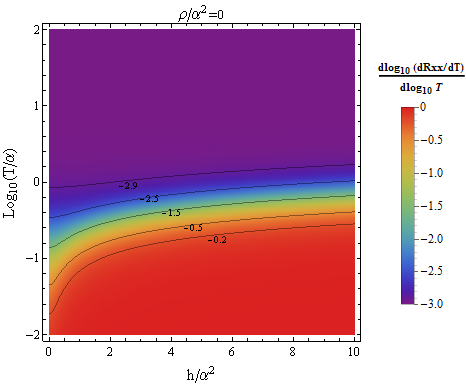}\includegraphics[scale=0.34]{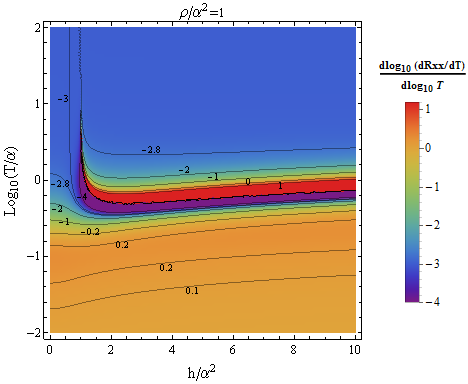}\includegraphics[scale=0.34]{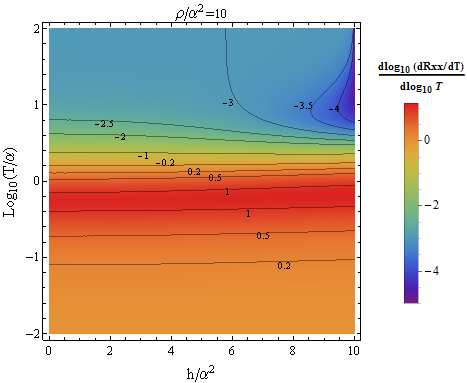}
\par\end{raggedright}
\caption{The temperature dependence of  $R_{xx}$ in  the Maxwell case. Upper row: Density plots of $d\log
_{10}(dR_{xx}/dT)/d\log_{10}T$ versus $\rho/\alpha^{2}$ and $\log
_{10}(T/\alpha)$ at fixed $h/\alpha^{2}=0$, $1$ and $10$ from left to right.
Lower row: Density plots of $d\log_{10}(dR_{xx}/dT)/d\log_{10}T$ versus
$h/\alpha^{2}$ and $\log_{10}(T/\alpha)$ at fixed $\rho/\alpha^{2}=0$, $1$,
and $10$ from left to right.}%
\label{figure Maxwell R}%
\end{figure}

We depict the density plots of $d\log_{10}(dR_{xx}/dT)/d\log_{10}T$ at some
fixed values of $h/\alpha^{2}$ or $\rho/\alpha^{2}$ in Figure
\ref{figure Maxwell R}. The vertical axis is labeled by $\log_{10}(T/\alpha)$,
and the range of the temperature $T/\alpha$ varies from $0.01$ to $100$.

From all figures in Figure \ref{figure Maxwell R}, we find two common
features. First, one finds that $N\sim-3$ at $T/\alpha\gtrsim10$. Actually, in
the high temperature limit $T/\alpha\gg(h/\alpha^{2},\rho/\alpha^{2})$, eqn.
(\ref{eq:Maxwell rh}) reduces to $T\propto r_{h}$, which leads to
\begin{equation}
R_{xx}\sim\text{constant}+(h^{2}-\rho^{2})(\alpha T)^{-2}+\mathcal{O}(T^{-4}).
\label{eq:high T limit}%
\end{equation}
More interestingly, we find that the resistivity varies linearly in
temperature at $T/\alpha\lesssim0.1$.

The density plots of $d\log_{10}(dR_{xx}/dT)/d\log_{10}T$ versus $\rho
/\alpha^{2}$ and $\log_{10}(T/\alpha)$ at $h/\alpha^{2}=0$, $1$ and $10$ are
displayed in the upper row of Figure \ref{figure Maxwell R}. At vanishing
magnetic field, as one increases the temperature, the corresponding $N$
monotonically decreases from $0$ to $-3$ at $\rho/\alpha^{2}\lesssim1$.
However at $\rho/\alpha^{2}\gtrsim1$, $N$ first increases from $0$ to a
maximum value and then decreases to $-3$. In the $h/\alpha^{2}=1$ case, the
scalings behavior at $\rho/\alpha\gtrsim1$ is similar to that in the previous
case while at $\rho/\alpha\lesssim1$ new behavior appears. One significant
character of this new behavior is the discontinuity between the purple region
and the red one as shown in the upper middle panel. The line separating these
two regions, we call it \textquotedblleft extremum line\textquotedblright, is
determined by $dR_{xx}/dT=0$ and thus the value of $d\log_{10}(dR_{xx}%
/dT)/d\log_{10}T$ diverges on this line resulting in the discontinuity.
Furthermore, $dR_{xx}/dT=0$ on the extremum line implies an extreme value of
$R_{xx}$ locally, indicating a transition between metal and insulator which is
consistent with \cite{Wang:2018hwg}. The presence of discontinuity provides
richer behavior and supports a wider spectrum of temperature scalings.
Increasing the magnetic field to $h/\alpha^{2}=10$, the extremum line
stretches across nearly all the values of $\rho/\alpha^{2}$ in the upper right
panel. The behavior below the extremum line is similar to the $h/\alpha^{2}=0$
case, but with much lower scalings exhibits. Due to the discontinuity, a
narrow strip-like region of T-linear resistivity survives into $T/\alpha
\gtrsim1$ above the extremum line.

The density plots of $d\log_{10}(dR_{xx}/dT)/d\log_{10}T$ versus $h/\alpha
^{2}$ and $\log_{10}(T/\alpha)$ at $\rho/\alpha^{2}=0$, $1$ and $10$ are
displayed in the lower row of Figure \ref{figure Maxwell R}. At vanishing
charge density, $N$ decreases monotonically from about $0$ to $-3$ as the
temperature increases. In the $\rho/\alpha^{2}=1$ case, the metal-insulator
transition appears and a narrow stripe of T-linear resistivity presents at
$T/\alpha\sim1$. For the case with $\rho/\alpha^{2}=10$, $N$ first increases
from $0$ to $1$ and then decreases to $-3$ with the increasing temperature.

To summarize, T-linear resistivity dominates in the low temperature regime
with $T/\alpha$$\lesssim0.1$ for almost all the range of $\rho/\alpha^{2}$ and
$h/\alpha^{2}$ in Figure \ref{figure Maxwell R} and survives into higher
temperatures in a narrow strip-like manner.

\subsubsection{Inverse Hall Angle}

\label{subsec:Maxwell Inverse-Hall-angle}

\begin{figure}[ptb]
\noindent\begin{raggedright}
\includegraphics[scale=0.34]{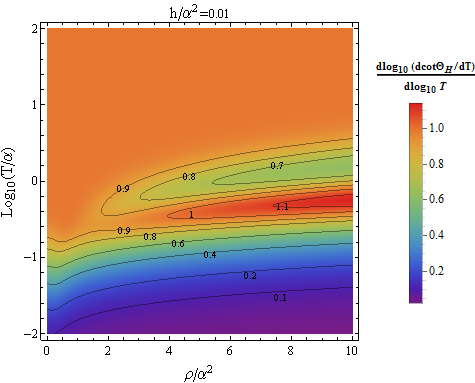}\includegraphics[scale=0.34]{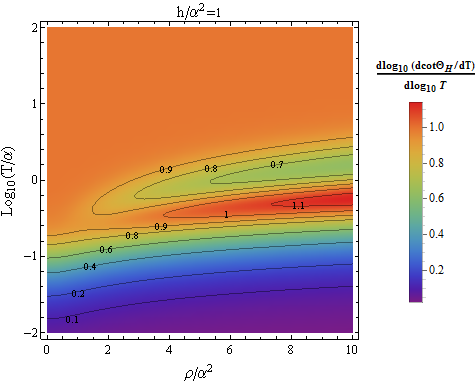}\includegraphics[scale=0.34]{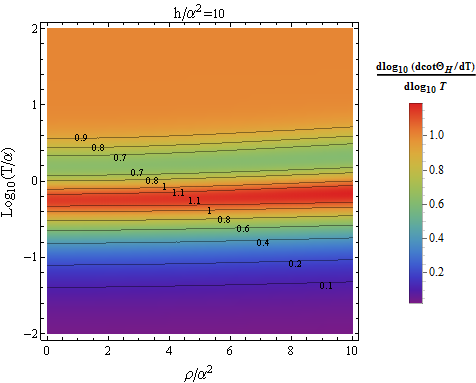}
\par\end{raggedright}
\caption{The temperature dependence of $\cot\Theta_{H}$ in the Maxwell case. Density plots of $d\log_{10}(d\cot\Theta_{H}/dT)/d\log_{10}T$ versus $\rho/\alpha^{2}$ and $\log_{10}(T/\alpha)$ at fixed $h/\alpha^{2}=0.01$, $1$ and $10$ from left to right. }%
\label{figure Maxwell C}%
\end{figure}

We display the density plots of $d\log_{10}(d\cot\Theta_{H}/dT)/d\log_{10}T$
at some fixed values of $h/\alpha^{2}$ in Figure \ref{figure Maxwell C}. Note
that $\cot\Theta_{H}$ remains invariant under the interchange between
$h/\alpha^{2}$ and $\rho/\alpha^{2}$ from eqns. $(\ref{eq:Maxwell C})$ and
$(\ref{eq:Maxwell rh}).$ Eqn. $\left(  \ref{eq:Maxwell C}\right)  $ shows that
$\cot\Theta_{H}$ diverges at $h/\alpha^{2}=0$, so we take a small but
non-vanishing magnetic field $h/\alpha^{2}=0.01$.

From Figure \ref{figure Maxwell C}, one can see that $M\sim1$ at
$T/\alpha\gtrsim10$, indicating the T-quadratic $\cot\Theta_{H}$ in the high
temperature regime. This is easy to understand from the high temperature limit
of eqn. $\left(  \ref{eq:Maxwell C}\right)  $. At $T/\alpha\lesssim0.1$, we
find that $M\sim0$. Similar to $R_{xx}$, the inverse Hall angle $\cot
\Theta_{H}$ behaves linearly in temperature at low temperatures. The two cases
with fixed $\rho/\alpha^{2}=0.01$ and $\rho/\alpha^{2}=1$ are similar. At
$h/\alpha^{2}\gtrsim4$ the scalings of these two cases both have a
non-monotonic behavior. As temperature increases, $M$ first increases from $0$
to about $1$, then decreases to about $0.7$, and then again increases to $1$.
In the $h/\alpha^{2}=10$ case, the pattern of the right panel of Figure
\ref{figure Maxwell C} is similar to those at $\rho/\alpha^{2}\gtrsim4$ of two
previous cases.

To summarize, T-quadratic $\cot\Theta_{H}$ not only dominates in the high
temperature regime but also extends to much lower temperatures, even
reaches $T/\alpha\sim0.1$ for small magnetic field and charge density.

\subsubsection{Overlap}

\label{subsec:Maxwell Overlap}

\begin{figure}[ptb]
\begin{centering}
\includegraphics[scale=0.34]{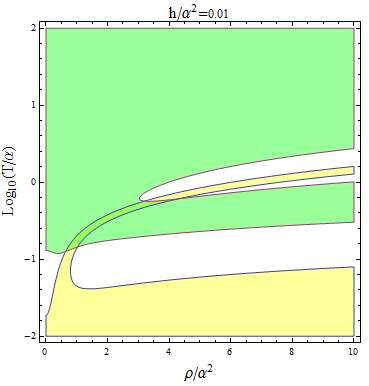}\includegraphics[scale=0.34]{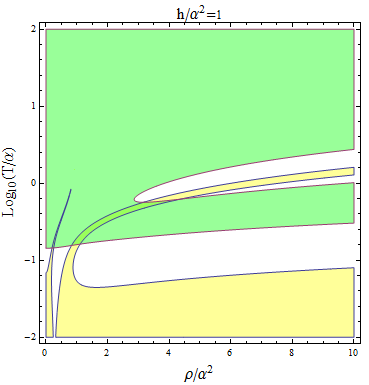}\includegraphics[scale=0.34]{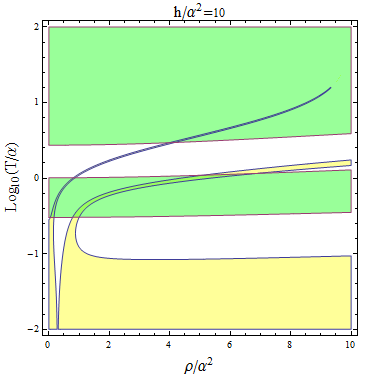}
\par\end{centering}
\begin{centering}
\includegraphics[scale=0.34]{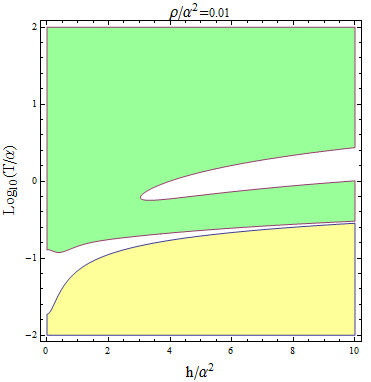}\includegraphics[scale=0.34]{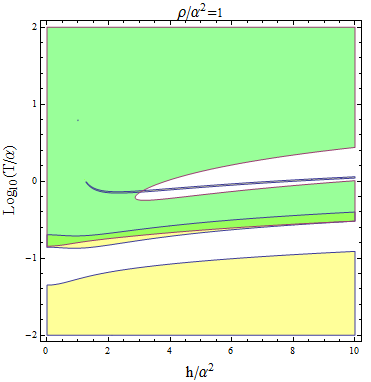}\includegraphics[scale=0.34]{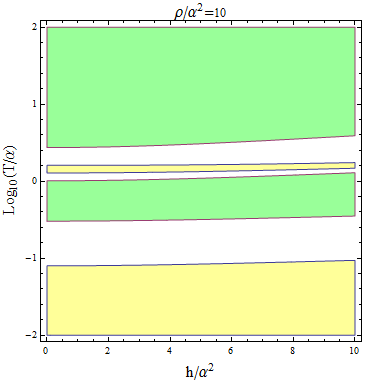}
\par\end{centering}
\begin{centering}
\includegraphics[scale=0.4]{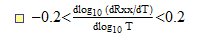}\includegraphics[scale=0.4]{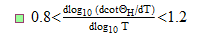}
\par\end{centering}
\caption{The overlap between T-linear $R_{xx}$ and T-quadratic $\cot\Theta_{H}$ in the Maxwell case. Upper Row: Region plots of $-0.2<d\log_{10}(dR_{xx}/dT)/d\log
_{10}T<0.2$ and $0.8<d\log_{10}(d\cot\Theta_{H}/dT)/d\log_{10}T<1.2$ versus
$\rho/\alpha^{2}$ and $\log_{10}(T/\alpha)$ at fixed $h/\alpha^{2}=0.01$, $1$
and $10$ from left to right. Lower Row: Region plots of $-0.2<d\log
_{10}(dR_{xx}/dT)/d\log_{10}T<0.2$ and $0.8<d\log_{10}(d\cot\Theta
_{H}/dT)/d\log_{10}T<1.2$ versus $h/\alpha^{2}$ and $\log_{10}(T/\alpha)$ at
fixed $\rho/\alpha^{2}=0.01$, $1$ and $10$ from left to right. The regions in
yellow and green correspond to the T-linear $R_{xx}$ and the T-quadratic
$\cot\Theta_{H}$, respectively. }%
\label{figure Maxwell overlap}%
\end{figure}

We check if there is overlap between the regions of T-linear $R_{xx}$ and
T-quadratic $\cot\Theta_{H}$. We approximately take $-0.2<N<0.2$ as the linear
temperature dependence of in-plane resistivity $R_{xx}$ and $0.8<M<1.2$ as the
quadratic temperature dependence of inverse Hall angle $\cot\Theta_{H}$, respectively.
In Figure \ref{figure Maxwell overlap}, we show the regions of T-linear
$R_{xx}$ and T-quadratic $\cot\Theta_{H}$ in yellow and green at some fixed
$h/\alpha^{2}$ or $\rho/\alpha^{2}$, respectively.

Generally speaking, the T-linear $R_{xx}$ dominates in the low temperature
regime with $T/\alpha\lesssim0.1$ and may survives into higher temperatures in
a narrow strip-like way, while T-quadratic $\cot\Theta_{H}$ dominates in the
high temperature regime with $T/\alpha\gtrsim10$ and can extend to lower
temperatures. The overlap does not occur in the cases with fixed $\rho
/\alpha^{2}=0.01$ and $10$. However as shown in Figure
\ref{figure Maxwell overlap}, there exists the overlap in other cases, which
occurs at $0.1\lesssim T/\alpha\lesssim1$.

\subsection{Born-Infeld Electrodynamics}

\label{subsec:Born-Infeld-Electrodynamics}

Born-Infeld theory is a typical nonlinear realization of electrodynamics
arising from the effective string theory at low energy scale with the Lagrange
given by
\begin{equation}
\mathcal{L}(s,p)=\frac{1}{a}(1-\sqrt{1-2as-a^{2}p^{2}}%
),\label{eq:BornInfeld Lagrange}%
\end{equation}
where the coupling parameter $a$ is related to the string tension
$\alpha^{\prime}$ as $a=(2\pi\alpha^{\prime})^{2}$. To get $R_{xx}$ and
$\cot\Theta_{H}$, we should first obtain $A_{t}^{^{\prime}}(r)$ in terms of
$r_{h}$, $a$, $h$ and $\rho$ by eqns. (\ref{eq:relations between}) and
(\ref{eq:BornInfeld Lagrange}):
\begin{equation}
A_{t}^{^{\prime}}(r_{h})=\frac{\rho}{\sqrt{r_{h}^{4}+a(h^{2}+\rho^{2})}%
}.\label{eq:At}%
\end{equation}
The reality of $A_{t}^{^{\prime}}(r_{h})$ gives a constraint
\begin{equation}
r_{h}^{4}+a(h^{2}+\rho^{2})>0.\label{eq:BornInfeld constraint}%
\end{equation}
For $a>0$, the above constraint holds automatically, while for $a<0$ it puts
an upper bound on the charge density and magnetic field. The physical
interpretation is that the singularity of the black brane needs to hide behind
the horizon. After some arrangements of eqns. $(\ref{eq:rh})$,
$(\ref{eq:relations between})$, $(\ref{eq:DC conductivities})$,
$(\ref{eq:resistivity and Hall angle})$, $(\ref{eq:BornInfeld Lagrange})$ and
$(\ref{eq:At})$, the resistivity $R_{xx}$ and inverse Hall angle $\cot
\Theta_{H}$ read
\begin{align}
R_{xx} &  =\frac{\alpha^{2}r_{h}^{2}\left(  h^{2}+\rho^{2}+\alpha^{2}%
\sqrt{r_{h}^{4}+a(h^{2}+\rho^{2})}\right)  }{\alpha^{4}r_{h}^{4}+\rho
^{2}\left(  a\alpha^{4}+h^{2}+\rho^{2}+2\alpha^{2}\sqrt{r_{h}^{4}+a(h^{2}%
+\rho^{2})}\right)  },\label{eq:BornInfeld resisitivity}\\
\cot\Theta_{H} &  =\frac{\alpha^{2}r_{h}^{2}\left(  h^{2}+\rho^{2}+\alpha
^{2}\sqrt{r_{h}^{4}+a(h^{2}+\rho^{2})}\right)  }{h\rho\left(  a\alpha
^{4}+h^{2}+\rho^{2}+2\alpha^{2}\sqrt{r_{h}^{4}+a(h^{2}+\rho^{2})}\right)
},\label{eq:BornInfeld Hall angle}%
\end{align}
with $r_{h}$ satisfying
\begin{equation}
-(1+6a)r_{h}^{2}+8\pi aTr_{h}+a\alpha^{2}+\sqrt{r_{h}^{4}+a(h^{2}+\rho^{2}%
)}=0.\label{eq:BornInfeld rh}%
\end{equation}
For $|a|\ll1$, Maxwell electrodynamics is recovered as expected. Note that
$\cot\Theta_{H}$ still possesses the symmetry between $h/\alpha^{2}$ and
$\rho/\alpha^{2}$.

\begin{figure}[ptb]
\noindent\begin{raggedright}
\includegraphics[scale=0.34]{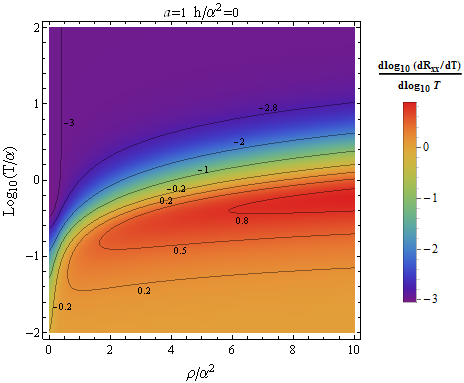}\includegraphics[scale=0.34]{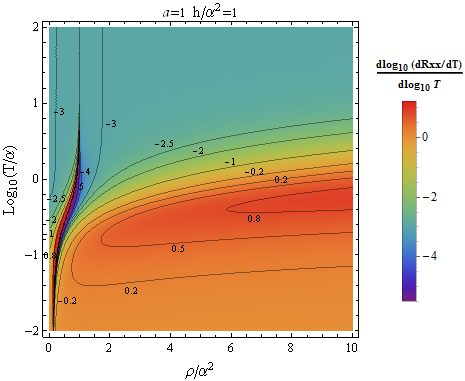}\includegraphics[scale=0.34]{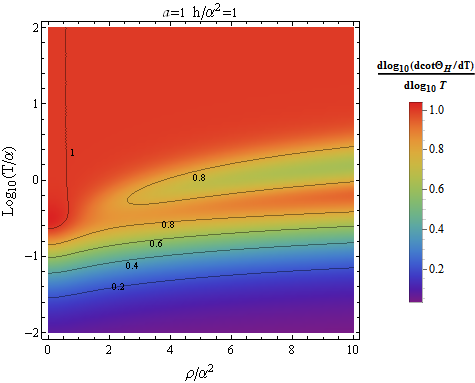}
\par\end{raggedright}
\noindent\begin{raggedright}
\includegraphics[scale=0.34]{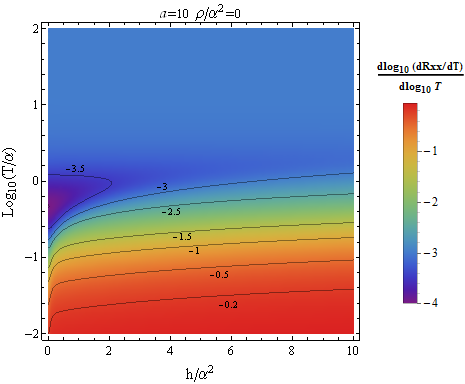}\includegraphics[scale=0.34]{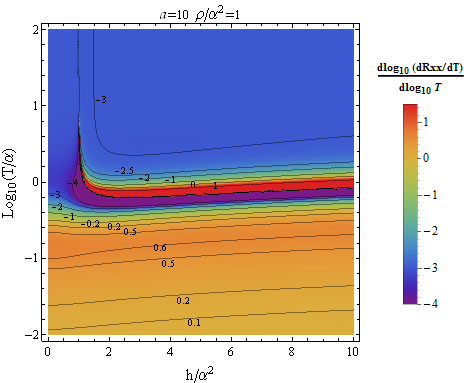}\includegraphics[scale=0.34]{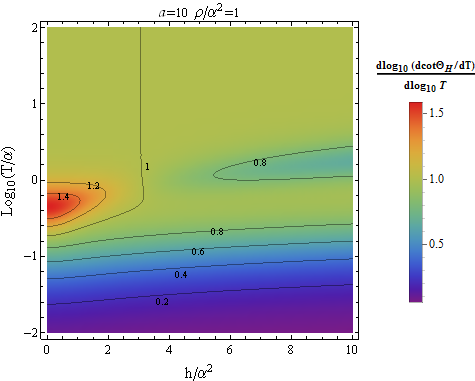}
\par\end{raggedright}
\caption{The temperature dependence of $R_{xx}$ and $\cot\Theta_{H}$ in the Born-Infeld case with $a>0$. Density plots of $d\log_{10}(dR_{xx}/dT)/d\log_{10}T$ and $d\log_{10}(d\cot\Theta_{H}/dT)/d\log_{10}T$ for various fixed values of $a$, $\rho
/\alpha^{2}$ and $h/\alpha^{2}$.}%
\label{figure BornInfeld a>0}
\end{figure}

At $a>0$, we find that the temperature dependence of $R_{xx}$ and $\cot
\Theta_{H}$ are quite similar to those of Maxwell electrodynamics so we only
show some examples in Figure \ref{figure BornInfeld a>0}. One could find that
the behavior in Figure \ref{figure BornInfeld a>0} is similar to that in
Figures \ref{figure Maxwell R} and \ref{figure Maxwell C}. The slight
difference between them is the range of $N$ and $M$. For instance, at
$\rho/\alpha^{2}=0$, the minimum of $N$ is $-3$ in the Maxwell case shown in
Figure \ref{figure Maxwell R} while it becomes $-3.5$ in the Born-Infeld case
shown in the lower left panel of Figure \ref{figure BornInfeld a>0}. The
similarity among the Maxwell case and Born-Infeld case with $a>0$ was also
noticed in \cite{Cremonini:2017qwq,Wang:2018hwg}.

Things become quite different in the $a<0$ case. In the following, we will
discuss the scalings of temperature dependence of $R_{xx}$ and $\cot\Theta
_{H}$ in various parameter spaces for $a<0$.

\subsubsection{In-plane Resistivity}

\label{subsec:BornInfeld In-plane-resistivity}

\begin{figure}[ptb]
\noindent
\includegraphics[scale=0.34]{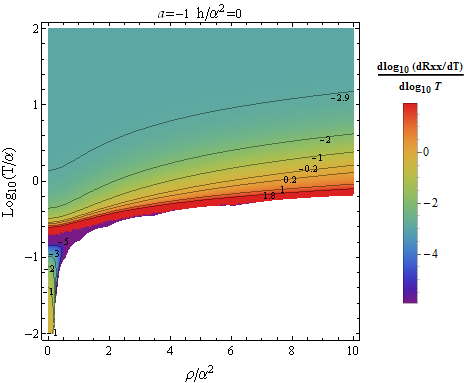}\includegraphics[scale=0.34]{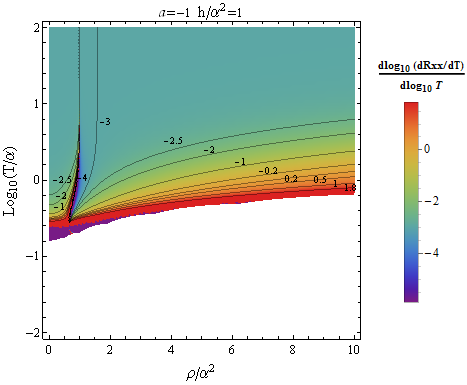}\includegraphics[scale=0.34]{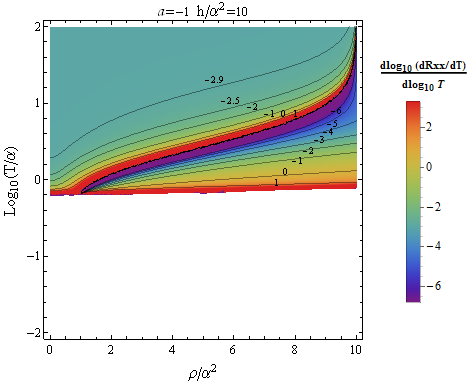}
\noindent\begin{raggedright}
\includegraphics[scale=0.34]{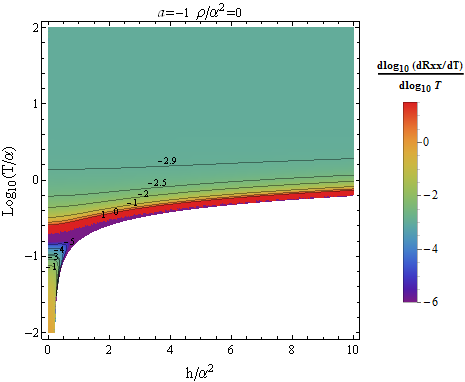}\includegraphics[scale=0.34]{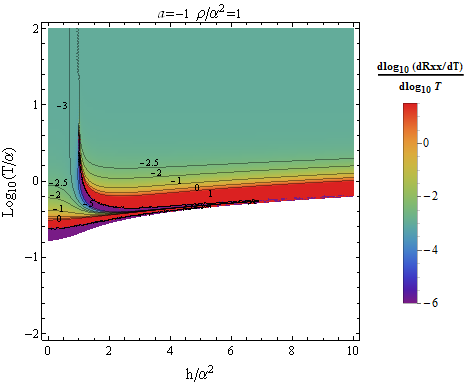}\includegraphics[scale=0.34]{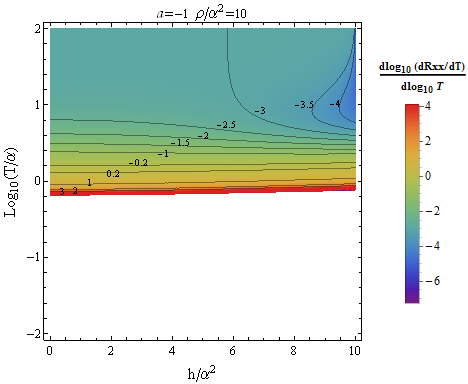}
\par\end{raggedright}
\caption{The temperature dependence of $R_{xx}$ in the Born-Infeld case with $a=-1$. Upper Row: Density plots of $d\log_{10}%
(dR_{xx}/dT)/d\log_{10}T$ versus $\rho/\alpha^{2}$ and $\log_{10}(T/\alpha)$
at fixed $h/\alpha^{2}=0$, $1$ and $10$ from left to right. Lower Row: Density
plots of $d\log_{10}(dR_{xx}/dT)/d\log_{10}T$ versus $h/\alpha^{2}$ and
$\log_{10}(T/\alpha)$ at fixed $\rho/\alpha^{2}=0$, $1$ and $10$ from left to
right.}%
\label{figure BornInfeld R}%
\end{figure}

In Figure \ref{figure BornInfeld R}, we depict the density plots of
$d\log_{10}(dR_{xx}/dT)/d\log_{10}T$ at fixed $a=-1$ and some values of
$h/\alpha^{2}$ or $\rho/\alpha^{2}$. We would only focus on $a=-1$ since the
cases with other fixed negative value of $a$ are much alike.

Compared to the case with $a>0$, the most obvious distinction is the
appearance of the white region due to the constraint
$(\ref{eq:BornInfeld constraint})$. The discontinuity usually occurs in the
region close to the white region, which implies the mental-insulator
transition. The behavior at $T/\alpha\gtrsim1$ in Figure
\ref{figure BornInfeld R} is reminiscent of that in Maxwell case. Though the
new behavior appears and the range of temperature scalings generally becomes
broader for $a<0$, the region of the T-liner resistivity becomes very smaller
than that in the $a>0$ case.

\begin{figure}[ptb]
\noindent\begin{raggedright}
\includegraphics[scale=0.34]{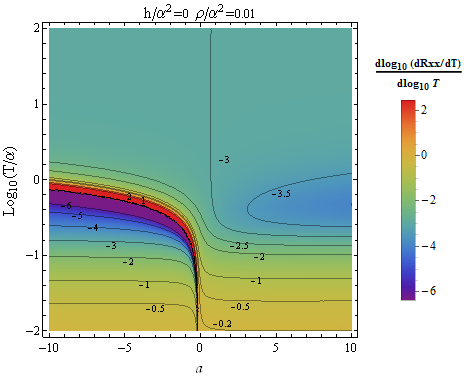}\includegraphics[scale=0.34]{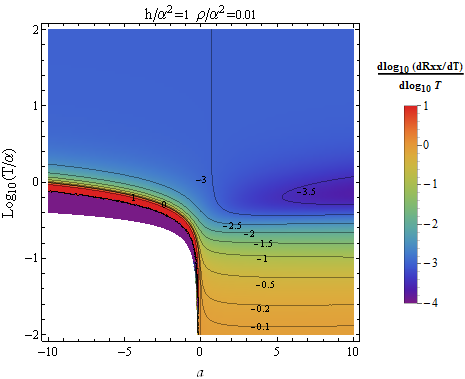}\includegraphics[scale=0.34]{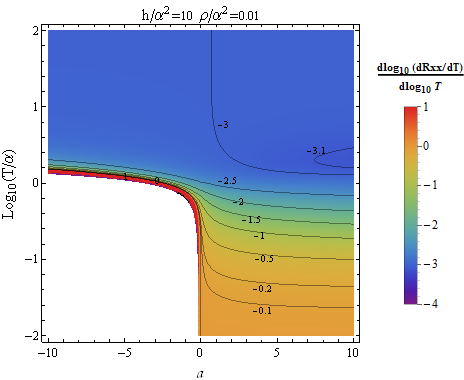}
\par\end{raggedright}
\noindent\begin{raggedright}
\includegraphics[scale=0.34]{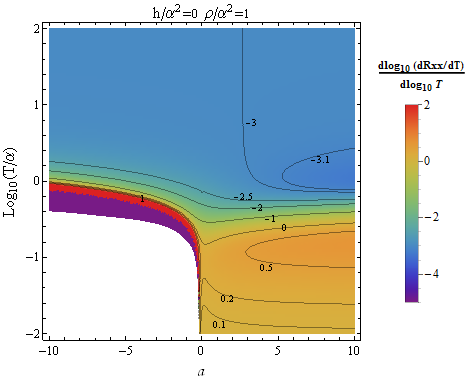}\includegraphics[scale=0.34]{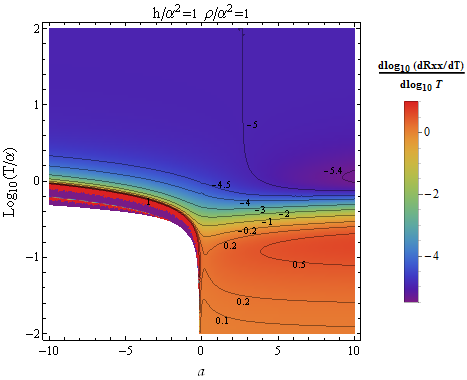}\includegraphics[scale=0.34]{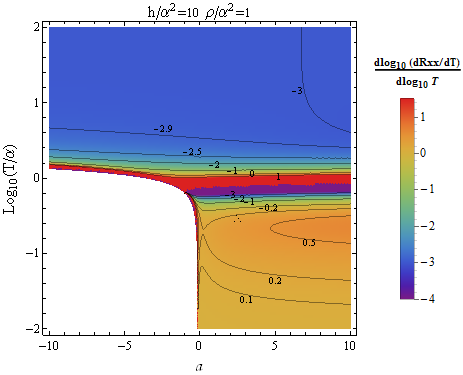}
\par\end{raggedright}
\noindent\begin{raggedright}
\includegraphics[scale=0.34]{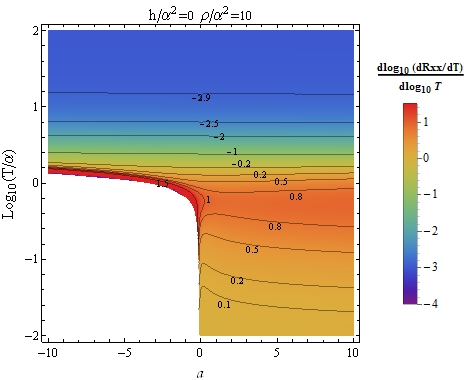}\includegraphics[scale=0.34]{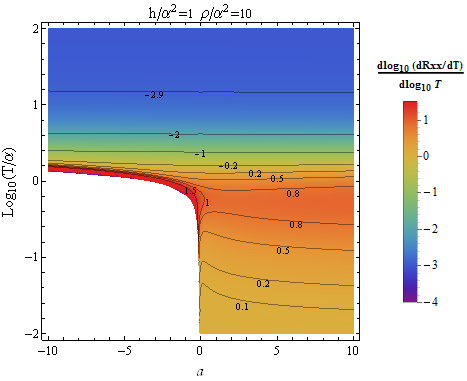}\includegraphics[scale=0.34]{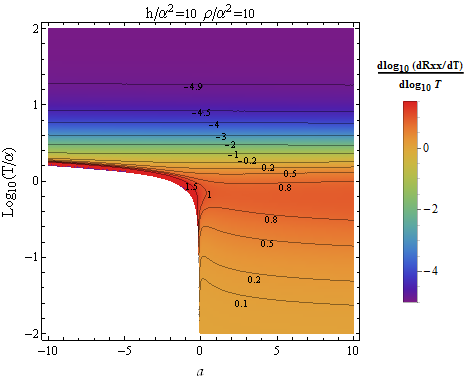}
\par\end{raggedright}
\caption{The temperature dependence of $R_{xx}$ in the Born-Infeld case. Density plots of $d\log_{10}(dR_{xx}/dT)/d\log_{10}T$ versus $a$ and
$\log_{10}(T/\alpha)$ at various fixed values of $h/\alpha^{2}$ and
$\rho/\alpha^{2}$. The fixed $\rho/\alpha^{2}$ for each row, from upper to
lower, are set as $0.01$, $1$ and $10$. And the fixed $h/\alpha^{2}$ for each
column, from left to right, are set as $0$, $1$ and $10$. }%
\label{figure BornInfeld h rho}%
\end{figure}

In the following, we study the temperature dependence of $R_{xx}$ with respect
to the parameter $a$ since it is a quantity characterizing the coupling of the
Born-Infeld electrodynamics. We show the density plots of $d\log_{10}%
(dR_{xx}/dT)/d\log_{10}T$ as a function of $a$ and $\log_{10}(T/\alpha)$ at
various fixed values of $h/\alpha^{2}$ and $\rho/\alpha^{2}$ in Figure
\ref{figure BornInfeld h rho}. Note that we choose a small but non-vanishing
charge density $\rho/\alpha^{2}=0.01$ in the upper row due to the triviality
of constant resistivity $R_{xx}=1$ for both strictly vanishing magnetic filed
and charge density.

We first focus on the upper row with fixed $\rho/\alpha^{2}=0.01$. In the
$h/\alpha^{2}=0$ case, there is no white region since $a$ is not large enough
to violate the constraint $(\ref{eq:BornInfeld constraint})$. At $a<0$, as the
temperature increases, $N$ first decreases from $0$ to $-\infty$, then
directly jumps to $+\infty$ on the extremum line and finally decreases to $-3$
at high temperatures as expected. And at $a>0$, as the temperature increases,
$N$ first decreases from $0$ to a minimum around $-3.5$ and then increases to
$-3$. The $h/\alpha^{2}=1$ and $10$ cases are quite similar to the
$h/\alpha^{2}=0$ case except the appearance of the white region.

We then consider the middle row with fixed $\rho/\alpha^{2}=1$. For vanishing
magnetic field, at $a>0$, $N$ first increases from $0$ to a maximum of around
$0.5$ and then decreases to $-3$ with the increasing temperature. For the
$h/\alpha^{2}=1$ case, $N\sim-5$ at sufficiently high temperatures rather than
the usual scaling $-3$. The reason is that in high temperature limit, the
$(T/\alpha)^{-2}$ term in eqn. $(\ref{eq:high T limit})$ vanishes due to
$h/\alpha^{2}=\rho/\alpha^{2}$, and thus the leading term dependent on the
temperature is at the order of $(T/\alpha)^{-4}$. Moreover, there are two
extremum lines at $a<0$, indicating two metal-insulator transitions at
$T/\alpha\sim1$. In the $h/\alpha^{2}=10$ case, in addition to the expected
discontinuity located at $a<0$, a new one appears at $a>0$.

We finally study the the lower row with fixed $\rho/\alpha^{2}=10$. In both
$h/\alpha^{2}=0$ and $h/\alpha^{2}=1$ cases, at $a>0$, $N$ first increases
from $0$ to a maximum of around $1$ and then decreases to $-3$ as the
temperature increases. In the case with $h/\alpha^{2}=10$, the reason for the
unusual scaling at high temperatures is same as that in the case with
$h/\alpha^{2}=1$ and $\rho/\alpha^{2}=1$.

Generally speaking, the regions of T-linear $R_{xx}$ in the Born-Infeld case
with $a>0$ is similar to those in the Maxwell case. For the Born-Infeld case
with $a<0$, they are strips at $T/\alpha\sim1$.

\subsubsection{Inverse Hall Angle}

\label{subsec:BornInfeld Inverse-Hall-angle}

\begin{figure}[ptb]
\noindent
\includegraphics[scale=0.34]{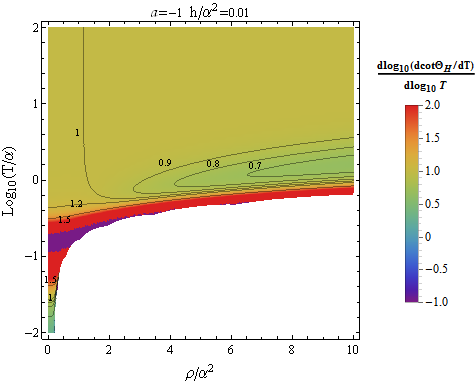}\includegraphics[scale=0.34]{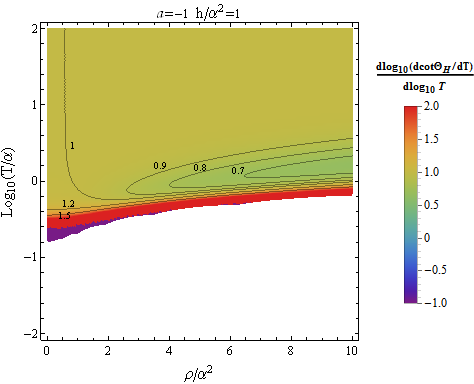}\includegraphics[scale=0.34]{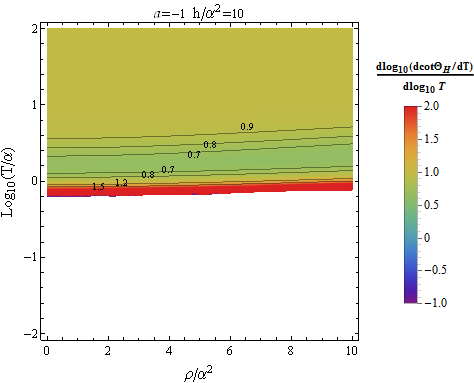}
\caption{The temperature dependence of $\cot\Theta_{H}$ in the Born-Infeld case with $a=-1$. Density
plots of $d\log_{10}(d\cot\Theta_{H}/dT)/d\log_{10}T$ versus $\rho/\alpha^{2}$
and $\log_{10}(T/\alpha)$ at fixed $h/\alpha^{2}=0.01$, $1$ and
$10$ from left to right.}%
\label{figure BornInfeld C h}%
\end{figure}

We now consider the scalings of temperature dependence of the inverse Hall
angle. We show the density plots of $d\log_{10}(d\cot\Theta_{H}/dT)/d\log
_{10}T$ at fixed $a=-1$ and some values of $h/\alpha^{2}$ in Figure
\ref{figure BornInfeld C h}. The cases at other fixed negative $a$ are in
quite similarity.

The behavior in the upper plane of the three figures in Figure
\ref{figure BornInfeld C h} is similar to that in the Maxwell case, which is
shown in Figure \ref{figure Maxwell C}. As before, the white region appears
due to the constraint (\ref{eq:BornInfeld constraint}). The presence of the
discontinuity largely widens the spectrum of the temperature scalings of
$\cot\Theta_{H}$. Note that a new type discontinuity here could come from the
vanishing denominator of $\cot\Theta_{H}$ $(\ref{eq:BornInfeld Hall angle})$.
Similar to the Maxwell and Born-Infeld with $a>0$ cases, the T-quadratic
$\cot\Theta_{H}$ dominates in the temperature regime with $T/\alpha\gtrsim1$.
In the case with $h/\alpha^{2}=0.01$, we find that T-quadratic $\cot\Theta
_{H}$ is observed at low temperatures with $T/\alpha\sim0.05$ for weak charge
density $\rho/\alpha^{2}\lesssim0.25$.

\begin{figure}[ptb]
\includegraphics[scale=0.34]{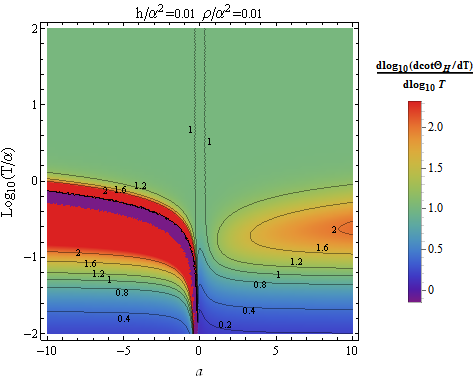}\includegraphics[scale=0.34]{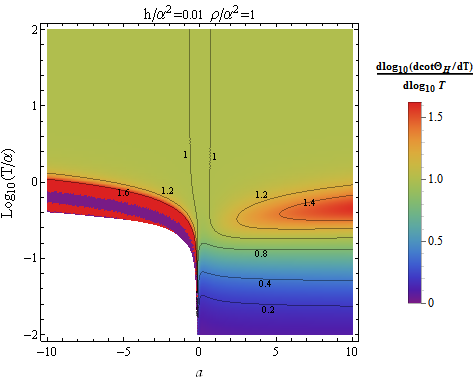}\includegraphics[scale=0.34]{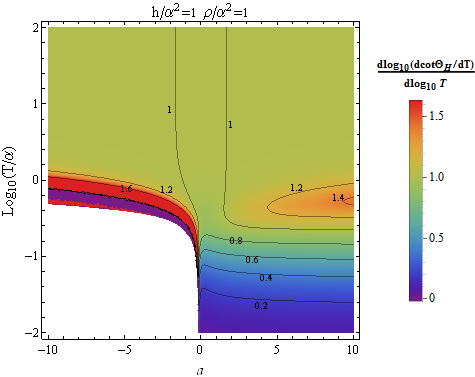}
\par
\includegraphics[scale=0.34]{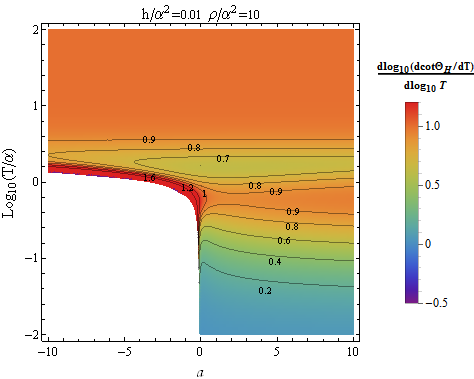}\includegraphics[scale=0.34]{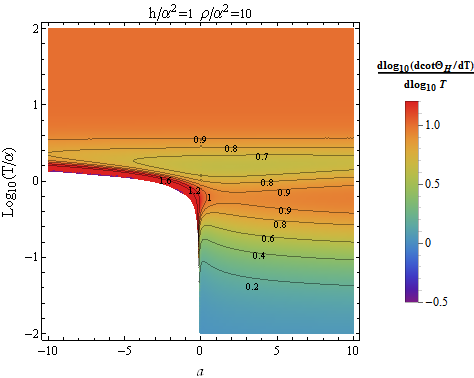}\includegraphics[scale=0.34]{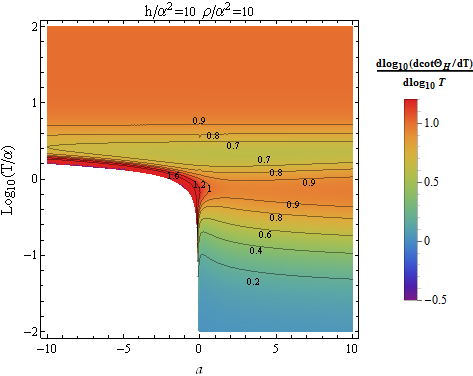}
\caption{The temperature dependence of $\cot\Theta_{H}$ in the Born-Infeld case. Density
plots of $d\log_{10}(d\cot\Theta_{H}/dT)/d\log_{10}T$ versus $a$ and
$\log_{10}(T/\alpha)$ at various fixed values of $h/\alpha^{2}$and
$\rho/\alpha^{2}$ after taking into account the symmetry between
$h/\alpha^{2}$ and $\rho/\alpha^{2}$.}%
\label{figure BornInfeld C h rho}%
\end{figure}

We display the density plots of $d\log_{10}(d\cot\Theta_{H}/dT)/d\log_{10}T$
against $a$ and $\log_{10}(T/\alpha)$ at various fixed values of $h/\alpha
^{2}$ and $\rho/\alpha^{2}$ in Figure \ref{figure BornInfeld C h rho}. For
small but non-vanishing magnetic filed and charge density as shown in the
upper left panel, no white region exhibits as expected. As the temperature
increases, $M$ increases monotonically from $0.4$ to $2$ in the region of
$a<0$ and $T/\alpha\lesssim0.1$. At $a>0$, $M$ first increases from $0.2$ to a
maximum and then decreases to $1$. T-quadratic $\cot\Theta_{H}$ dominates in
$T/\alpha\gtrsim1$ for all values of $a$, and also presents at $T/\alpha
\sim0.05$ for $a<0$, which is consistent with the $h/\alpha^{2}=0.01$ case in
Figure \ref{figure BornInfeld C h}. For the cases in the upper middle and
upper right panels, both are similar to the previous case except the presence
of the white region. The three cases in lower row are all similar. At $a>0$,
with the increasing temperature, $M$ first increases from $0$ to around $0.9$,
then decreases to around $0.6$ and finally increases to $1$. The region of
T-quadratic $\cot\Theta_{H}$ presents in $T/\alpha\gtrsim1.5$ and
$0.25\lesssim T/\alpha\lesssim1$.

\subsubsection{Overlap}

\label{subsec:BornInfeld overlap}

\begin{figure}[ptb]
\noindent\begin{centering}
\includegraphics[scale=0.35]{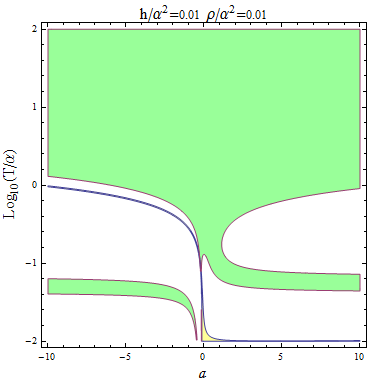}\includegraphics[scale=0.35]{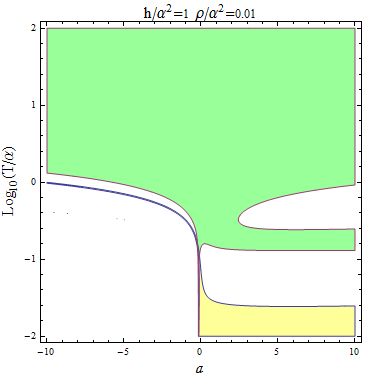}\includegraphics[scale=0.35]{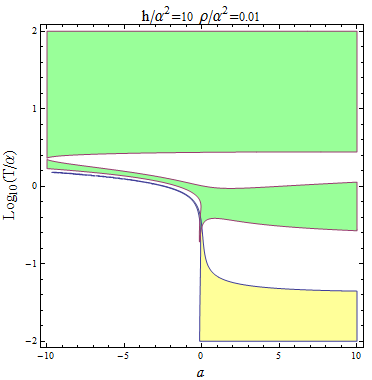}
\par\end{centering}
\noindent\begin{centering}
\includegraphics[scale=0.35]{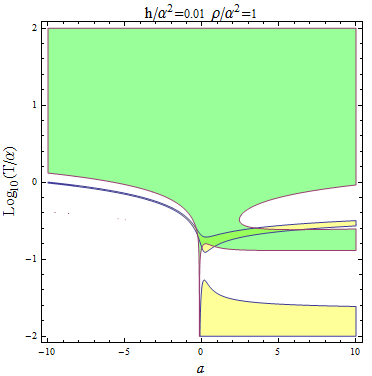}\includegraphics[scale=0.35]{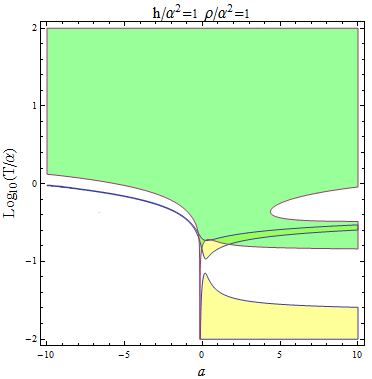}\includegraphics[scale=0.35]{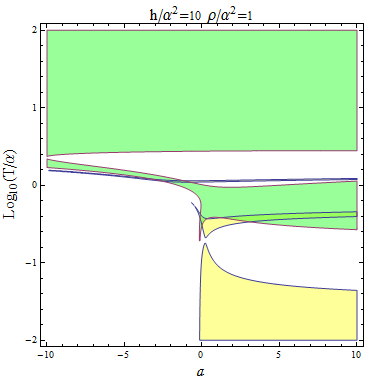}
\par\end{centering}
\noindent\begin{centering}
\includegraphics[scale=0.35]{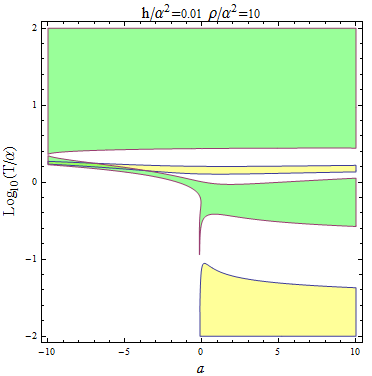}\includegraphics[scale=0.35]{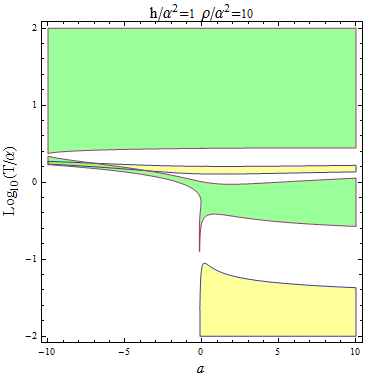}\includegraphics[scale=0.35]{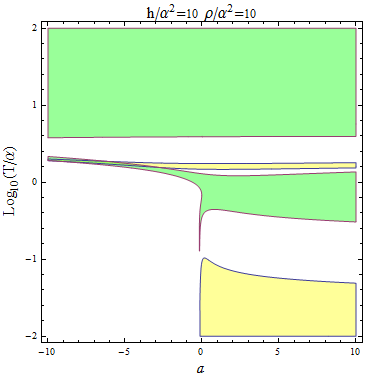}
\par\end{centering}
\begin{centering}
\includegraphics[scale=0.4]{m0202.png}\includegraphics[scale=0.4]{0812.png}
\par\end{centering}
\caption{The overlap between T-linear $R_{xx}$ and T-quadratic $\cot\Theta_{H}$ in the Born-Infeld case. Region plots of $-0.2<d\log_{10}(dR_{xx}/dT)/d\log_{10}T<0.2$ and
$0.8<d\log_{10}(d\cot\Theta_{H}/dT)/d\log_{10}T<1.2$ versus $a$ and $\log
_{10}(T/\alpha)$ for several fixed values of $h/\alpha^{2}$ and $\rho
/\alpha^{2}$. The fixed $\rho/\alpha^{2}$ for each row, from upper to lower,
are set as $0.01$, $1$ and $10$. And the fixed $h/\alpha^{2}$ for each column,
from left to right, are set as $0.01$, $1$ and $10$. The regions in yellow and
green correspond to the T-linear $R_{xx}$ and the T-quadratic $\cot\Theta_{H}%
$, respectively. }%
\label{figure BornInfeld overlap}%
\end{figure}

We end this section by discussing the overlap between T-linear $R_{xx}$ and
T-quadratic $\cot\Theta_{H}$ for Born-Infeld electrodynamics. For the $a>0$
case, the overlaps plotted in the $h/\alpha^{2}$ $(\rho/\alpha^{2})$%
-$\log_{10}\left(  T/\alpha\right)  $ plane with fixed $a$ are similar to
those in the Maxwell cases in Figure \ref{figure Maxwell overlap}. For the
$a<0$ case, the region of T-linear $R_{xx}$ in the $h/\alpha^{2}$
$(\rho/\alpha^{2})$-$\log_{10}\left(  T/\alpha\right)  $ plane with fixed $a$
is very small as shown in Figures \ref{figure BornInfeld R}. In Figure
\ref{figure BornInfeld overlap}, we display the region plots of T-linear
$R_{xx}$ in yellow and T-quadratic $\cot\Theta_{H}$ in green as a function of
$a$ and $\log_{10}(T/\alpha)$ at several fixed values of $h/\alpha^{2}$ and
$\rho/\alpha^{2}$.

Generally speaking, T-linear $R_{xx}$ mainly lives in low temperatures with
$T/\alpha\lesssim0.1$ for $a>0$ and survives in some strip-like regions at
$T/\alpha\gtrsim0.1$ for all values of $a$. And T-quadratic $\cot\Theta_{H}$
dominates at high temperatures for all range of $a$ and extends down to low
temperatures. In the upper row, no overlap exhibits. In the middle row with
fixed $\rho/\alpha^{2}=1$, the overlap would occur at $a>0$ and $0.1\lesssim
T/\alpha\lesssim1$, which is reminiscent of the Maxwell case in Figure
\ref{figure Maxwell overlap}. In the third row with $\rho/\alpha^{2}=10$, a
narrow strip-like overlap presents at $a\lesssim-5$ and $T/\alpha\sim1$,
distinguishing from the Maxwell case.

\section{Discussion and Conclusion}

\label{sec:Discussion-and-conclusion}

In this paper, we investigated the temperature dependence of the in-plane
resistivity $R_{xx}$ and inverse Hall angle $\cot\Theta_{H}$ for the NLED
holographic model developed in our previous work \cite{Wang:2018hwg}. To
extract the effective scalings of temperature dependence of $R_{xx}$ and
$\cot\Theta_{H}$, we took the advantage of the density plot of $d\log
_{10}(dR_{xx}/dT)/d\log_{10}T$ and $d\log_{10}(d\cot\Theta_{H}/dT)/d\log
_{10}T$ in parameter space. In section \ref{sec:In-Plane-transport-properties}%
, we focused on two specific cases in the model: one is Maxwell
electrodynamics and the other is the nonlinear Born-Infeld electrodynamics.

For Maxwell electrodynamics, a wide spectrum of the scalings of the
temperature dependence of $R_{xx}$ and $\cot\Theta_{H}$ can be observed. In
general, the in-plane resistivity has been shown to vary as $R_{xx}\sim T$ at
low temperatures and $R_{xx}\sim T^{-2}$ at high temperatures. And the inverse
Hall angle varies as $\cot\Theta_{H}\sim T$ at low temperatures and
$\cot\Theta_{H}\sim T^{2}$ at high temperatures. Moreover, the presence of
discontinuity in the density plot of $d\log_{10}(dR_{xx}/dT)/d\log_{10}T$
implies the metal-insulator transition. In general, the T-liner $R_{xx}$
dominates at low temperatures with $T/\alpha\lesssim0.1$ and might survives
into higher temperatures in a narrow strip-like manner. And the T-quadratic
$\cot\Theta_{H}$ dominates at high temperatures with $T/\alpha\gtrsim10$ and
extends down to lower temperatures, even to $T/\alpha\sim0.1$ at small
 magnetic field and charge density. The overlap, if occurs,
generally locates in the intermediate temperate regime within $0.1\lesssim
T/\alpha\lesssim1$ as shown in Figure \ref{figure Maxwell overlap}.

For nonlinear Born-Infeld electrodynamics with $a>0$, the temperature
dependence of $R_{xx}$ and $\cot\Theta_{H}$ and their overlap are quite
similar to Maxwell case. While at $a<0$, the constraint
$(\ref{eq:BornInfeld constraint})$ generally results in white region at low
temperatures, which provides richer behavior and broader range of scalings
than Maxwell case. At $a<0$, the T-linear $R_{xx}$ presents in a narrow
strip-like region at $T/\alpha\sim1$ and T-quadratic $\cot\Theta_{H}$ still
dominates at high temperatures. And the overlap in the $a<0$ case could occur
at strong charge density as shown in Figure \ref{figure BornInfeld overlap}.

Notice that we used rescaled quantities, e.g., the rescaled temperature, all
the time. However, it is quite unknown how the value of $\alpha$ is related to
experiments. Though the area of the overlap seems small in both
electrodynamics, it might explain T-liner $R_{xx}$ and T-quadratic $\cot
\Theta_{H}$ simitaneously appearing in experiments for some suitable value of
$\alpha$. Furthermore, increasing the degrees of freedom in holographic models
could enlarge the overlap. For instance, one could introduce the dilaton field
which is widely investigated in literatures, e.g., \cite{Mike Blake,Erin
Blauvelt,Cremonini:2017qwq}.{}

Finally, we found that in the Maxwell case with vanishing magnetic field, see
Figure \ref{figure Maxwell R}, $N$ ranges from $0$ to $1$ at $T/\alpha
\lesssim1$. It might imply a combination dependence of temperature $T+T^{2}$
on $R_{xx}$, which could help to explain the unconventional behaviors observed
at low temperatures of two prototypical copper oxide superconductors LSCO and
TBCO \cite{CooperWang}. It deserves future study to make concrete comparisons
with experiments.

\vspace*{4ex}
\noindent\textbf{Acknowledgements}

We are grateful to Shuxuan Ying and Houwen Wu for useful discussions and
valuable comments. This work is supported in part by NSFC (Grant No. 11005016,
11175039 and 11375121).

\end{document}